\documentclass[twocolumn,preprintnumbers,amsmath,amssymb]{revtex4}

\usepackage{amsmath}
\usepackage{amsfonts}
\usepackage{amssymb}
\usepackage{graphicx}
\usepackage{dcolumn}
\usepackage{bm}
\usepackage{xspace}
\usepackage{color}
\usepackage{calrsfs}
\DeclareMathAlphabet{\mathpzc}{OT1}{pzc}{m}{it}

\newcommand{\sqrts}{\sqrt{s}}

\newcommand{\pp}{{\rm{pp}}}

\newcommand{\ppbar}{{\rm{p$\bar{\rm p}$}}}

\newcommand{\pythia}{{\sc pythia}}
\newcommand{\herwig}{{\sc herwig}}

\newcommand{\sigmaSPS}{\sigma^{{\rm {\tiny SPS}}}}

\newcommand{\sigmaTPS}{\sigma^{{\rm {\tiny TPS}}}}
\newcommand{\sigmaeff}{\sigma_{\rm eff,DPS}}
\newcommand{\sigmaefftps}{\sigma_{\rm eff,TPS}}

\newcommand{\QQbar}    {\rm Q\overline{Q}}
\providecommand{\ccbar}{\rm c\overline{c}}
\providecommand{\bbbar}{\rm b\overline{b}}

\newcommand*{\ie}{i.e.\@\xspace}
\newcommand*{\cm}{c.m.\@\xspace}

\def\ttt#1{\texttt{\small #1}}
\newcommand{\toppp}{\ttt{Top++}}

\begin{document}

\title{Triple parton scatterings in high-energy proton-proton collisions}

\author{David d'Enterria}
\affiliation{CERN, EP Department, 1211 Geneva, Switzerland}
\author{\vspace{-0.3cm} Alexander M.~Snigirev}
\affiliation{Skobeltsyn Institute of Nuclear Physics, Lomonosov Moscow State University, 119991, Moscow, Russia}

\date{\today}

\begin{abstract}
A generic expression to compute triple parton scattering (TPS) cross sections in high-energy proton-proton 
(\pp) collisions is presented as a function of the corresponding single-parton cross sections and the transverse 
parton distribution in the proton encoded in an effective parameter $\sigmaefftps$. The value of $\sigmaefftps$ 
is closely related to the similar effective cross section that characterizes double-parton scatterings, 
and amounts to $\sigmaefftps = 12.5 \pm 4.5$~mb. Estimates for triple charm ($\ccbar$) and bottom ($\bbbar$) 
production in \pp\ collisions at LHC and FCC energies are presented based on next-to-next-to-leading order 
perturbative calculations for single $\ccbar,\bbbar$ cross sections. At $\sqrts\approx$~100~TeV, about 15\%
of the \pp\ collisions produce three $\ccbar$ pairs from three different parton-parton scatterings.
\end{abstract}


\maketitle

The existence of multiparton interactions (MPI) in high-energy hadronic collisions is a natural consequence of the 
finite size of hadrons, the fast increase of the parton flux at small parton longitudinal momentum fractions (Bjorken $x$),
and the requirement of unitarization of the cross sections in perturbative 
quantum chromodynamics (pQCD)~\cite{Diehl:2011yj}. The field of MPI has attracted an increasing interest in the last 
years motivated, among other reasons, by the accumulated experimental evidence of double parton scattering (DPS) 
processes concurrently producing two independently-identified hard particles in the same proton-(anti)proton 
(\pp, \ppbar) collision, 
at Tevatron and LHC energies~\cite{Bartalini:2011jp,Abramowicz:2013iva,Bansal:2014paa,Astalos:2015ivw}. 
Multiple hard parton interaction rates depend chiefly on the degree of overlap between the matter distributions 
of the colliding hadrons~\cite{dEnterria:2010xip}, 
and provide valuable information on the poorly known transverse parton profile of the proton, the unknown energy 
evolution of the parton density as a function of impact parameter ($b$), as well as on the role of many-parton 
correlations in the hadronic wave functions~\cite{Calucci:2010wg}.

In this paper, we present for the first time a quantitative estimate 
of the cross section for observing three separate hard interactions in a \pp\ 
collision (triple parton scattering, TPS), a possibility considered earlier~\cite{CalucciTreleani,Maina:2009sj,Snigirev:2016uaq},
but for which no simple formula of the expected rates based on the underlying single-parton scattering (SPS) 
cross sections 
existed so far. 
A good understanding of TPS is not only useful to improve our knowledge of the 3D parton structure of the proton,
but is also of relevance for a realistic characterization of backgrounds for rare final-states with multiple 
heavy-particles in searches of new physics.
As we show below, TPS can represent a non-negligible fraction of some particular final states at future hadron 
colliders such as FCC, projected to deliver several 100~fb$^{-1}$/year integrated luminosities 
in \pp\ collisions at a \cm\ energy of $\sqrts$~=~100~TeV~\cite{Mangano16b}.

We review first the theoretical expression for TPS cross sections in a generic hadron-hadron collision,
expressed as a convolution of SPS cross sections and generalized parton densities (dependent on $x$, $Q^2$ and $b$), 
and derive a simple factorized expression for the TPS cross sections as a triple product of SPS cross sections
normalized by an effective cross section $\sigmaefftps$ characterizing the transverse area 
of triple partonic interactions. From simple geometric considerations, we show how $\sigmaefftps$ is 
closely connected to the effective cross section $\sigmaeff$ already measured in DPS.
From the existing measurements of $\sigmaeff$ and realistic proton profiles, 
we derive the numerical value of $\sigmaefftps$, which proves very robust 
with respect to any particular choice of the underlying transverse parton density. As a concrete numerical example,
we provide estimates for triple charm ($\ccbar$) and bottom ($\bbbar$) cross sections in \pp\ collisions at
the LHC and FCC, based on next-to-next-to-leading-order (NNLO) calculations
of the corresponding SPS cross sections. 


In a generic hadron-hadron collision, the inclusive TPS cross section from three independent 
parton subprocesses ($h h' \to abc$)  can be written~\cite{CalucciTreleani,Maina:2009sj,Snigirev:2016uaq} 
as a convolution of generalized parton distribution functions (PDFs) and elementary cross sections summed over all involved partons
\begin{eqnarray} 
\label{hardAB}
& &\sigma^{\rm TPS}_{hh' \to abc} \nonumber\\  
& & = \frac{\mathpzc{m}}{3!} \sum \limits_{i,j,k,l,m,n} \int \Gamma^{ijk}_{h}(x_1, x_2, x_3; {\bf b_1},{\bf b_2}, {\bf b_3}; Q^2_1, Q^2_2, Q^2_3)\nonumber\\
& &\times\hat{\sigma}_a^{il}(x_1, x_1^{'},Q^2_1) \hat{\sigma}_b^{jm}(x_2, x_2^{'},Q^2_2)\hat{\sigma}_c^{kn}(x_3, x_3^{'},Q^2_3)\nonumber\\
& &\times\Gamma^{lmn}_{h'}(x_1^{'}, x_2^{'}, x_3^{'}; {\bf b_1} - {\bf b},{\bf b_2} - {\bf b},{\bf b_3} - {\bf b}; Q^2_1, Q^2_2, Q^2_3)\nonumber\\
& &\times dx_1 dx_2 dx_3 dx_1^{'} dx_2^{'} dx_3^{'} d^2b_1 d^2b_2 d^2b_3 d^2b.
\end{eqnarray}
In this expression, $\Gamma^{ijk}_{h}(x_1, x_2, x_3; {\bf b_1},{\bf b_2}, {\bf b_3}; Q^2_1, Q^2_2, Q^2_3)$ 
are the triple parton distribution functions, depending on the momentum fractions $x_1$, $x_2$, $x_3$ at 
transverse positions ${\bf b_1}$, ${\bf b_2}$, ${\bf b_3}$ of the three partons $i$, $j$, $k$, 
producing final states $a$, $b$, $c$ at energy scales $Q_1$, $Q_2$, $Q_3$, with
subprocess cross sections $\hat{\sigma}_a^{il}$, $\hat{\sigma}_b^{jm}$, $\hat{\sigma}_c^{kn}$. 
The combinatorial prefactor $\mathpzc{m}/3!$ takes into the different cases of 
(indistinguishable or not) final states: $\mathpzc{m}=1$ if $a=b=c$; $\mathpzc{m}=3$ if $a=b$, 
or $a=c$, or $b=c$; and  $\mathpzc{m}=6$ if $a, b, c$ are different.
The triple parton distribution functions 
$\Gamma^{ijk}_{h}(x_1, x_2, x_3; {\bf b_1},{\bf b_2}, {\bf b_3}; Q^2_1, Q^2_2, Q^2_3)$ encode all the 
parton structure information of relevance for TPS, and are typically assumed to be decomposable in terms 
of longitudinal and transverse components
\begin{eqnarray} 
\label{DxF}
& &\Gamma^{ijk}_{h}(x_1, x_2, x_3; {\bf b_1},{\bf b_2}, {\bf b_3}; Q^2_1, Q^2_2, Q^2_3)\nonumber\\
& &= D^{ijk}_h(x_1, x_2, x_3; Q^2_1, Q^2_2, Q^2_3) f({\bf b_1}) f({\bf b_2}) f({\bf b_3}),
\end{eqnarray} 
where $f({\bf b_1})$ describes the transverse parton density, often assumed to be a universal function 
for all types of partons, from which the corresponding hadron-hadron overlap function is derived:
\begin{equation} 
\label{f}
\mkern-14mu T({\bf b}) = \int f({\bf b_1}) f({\bf b_1 -b})d^2b_1 ,\,{\rm with}\,\int d^2b T({\bf b})=1.
\end{equation} 
Making the further assumption that the longitudinal components
reduce to the product of independent single PDFs, 
$D^{ijk}_h(x_1, x_2, x_3; Q^2_1, Q^2_2, Q^2_3) = D^i_h (x_1; Q^2_1) D^j_h (x_2; Q^2_2) D^k_h (x_3; Q^2_3)$,
the cross section of TPS can be expressed in the simple generic form
\begin{equation} 
\label{doubleAB}
\sigma_{hh' \to abc  }^{\rm TPS} =  \left(\frac{\mathpzc{m}}{3!}\right)\, \frac{\sigma_{hh' \to a}^{\rm SPS} \cdot
\sigma_{ hh' \to b}^{\rm SPS} \cdot \sigma_{ hh' \to c}^{\rm SPS}}{\sigmaefftps^2},
\end{equation} 
\ie, as a triple product of independent single inclusive cross sections
\begin{eqnarray} 
\sigmaSPS_{hh' \to a} =  \sum \limits_{i,k} \int D^{i}_h(x_1; Q^2_1) f({\bf b_1})\, \hat{\sigma}^{ik}_{a}(x_1, x_1') \nonumber\\
 \times\, D^{k}_{h'}(x_1'; Q^2_1)f( {\bf b_1} - {\bf b}) dx_1 dx_1' d^2b_1 d^2b  \nonumber\\
= \sum \limits_{i,k} \int D^{i}_h(x_1; Q^2_1) \,\hat{\sigma}^{ik}_{a}(x_1, x_1') \,D^{k}_{h'}(x_1'; Q^2_1) dx_1 dx_1',\label{eq:hardS}
\end{eqnarray}
normalized by the square of an effective TPS cross section
\begin{eqnarray} 
\sigmaefftps^2=\left[ \int d^2b \,T^3({\bf b})\right]^{-1}\,,
\label{eq:sigmaeffTPS}
\end{eqnarray} 
which is closely related to the similar quantity
\begin{eqnarray} 
\label{eq:sigmaeff1}
\sigmaeff=\left[ \int d^2b\,T^2({\bf b})\right]^{-1} \,,
\end{eqnarray}
determined to be $\sigmaeff \simeq 15\pm5$~mb in DPS measurements at Tevatron and the LHC~\cite{Astalos:2015ivw,Abe:1997xk,Khachatryan:2015pea,Aaboud:2016dea,Aaij:2015wpa}.

Whereas any TPS cross section can {\it always} be expressed as the triple product~(\ref{doubleAB}),
with $\sigmaefftps$ encoding all unknowns about the TPS dynamics, the geometrical
interpretation of this latter quantity given by Eq.~(\ref{eq:sigmaeffTPS}) relies on
the assumption of the aforementioned simplifying yet economical assumptions on the
factorization of longitudinal and transverse degrees of freedom, the absence of 
multiparton correlations, and flavor (gluon, quark)-independent transverse parton profiles. 
In addition, there is no a priori reason to take the transverse hadron profile $f({\bf b})$, 
and therefore also $\sigmaefftps$ and $\sigmaeff$, as a constant with collision energy.
As a matter of fact, DPS studies~\cite{Abe:1997xk,Seymour:2013qka,Khachatryan:2015pea} indicate that the experimentally-extracted 
$\sigmaeff\simeq$~15~mb values are about a factor of two smaller (\ie, the DPS cross sections are 
about twice larger) than expected from Eq.~(\ref{eq:sigmaeff1}) for a ``standard'' proton geometric profile. 
The concurrent measurement of $\sigmaefftps$ and $\sigmaeff$ in different colliding systems and for 
different final-states can help clarify all these open issues, shedding light on 
the 3D partonic structure of the proton and its evolution as a function of energy.

To estimate the value of $\sigmaefftps$, let's consider first a simplistic class of analytical 
models for the overlap function of a given hadronic AA collision
\begin{eqnarray} 
T_{\rm AA}({\bf b}) = \left\{ \begin{array}{@{}ll@{}}
\frac{\rm A^2 (n+1)}{4\pi R^2_A} [1-(b/(2R_{\rm A}))^2]^n &,\; b<2R_{\rm A}, \\
0 &,\; b>2R_{\rm A} 
\end{array}\right.
\label{h-s,T}
\end{eqnarray}
normalized to $\int d^2 b~ T_{\rm AA}({\bf b})= \rm A^2$, where exponents $n=0,1,\infty$ give respectively
a flat, hard-sphere-like, and Gaussian-like distribution 
in the interval [0, $2\,R_A$].  
This generic expression applies to any hadronic system, including 
collisions of nuclei with nucleon number A~\cite{dEnterriaSnigirev}, 
but we will mostly focus below on the proton case for which $\rm A=1$, and 
$R_{\rm A}=r_p$ is its characteristic transverse ``radius''. From Eq.~(\ref{eq:sigmaeffTPS}) one obtains:
\begin{eqnarray} 
\sigmaefftps^2=\left[ \int d^2 b~ T^3_{\rm AA}({\bf b})\right]^{-1}= \frac{(4\pi R^2_A)^2 (3n+1)}{\rm A^6 (n+1)^3}.
\label{h-s,T3}
\end{eqnarray}
Instead of an expression which depends on $R_{\rm A}$, it's more convenient to express the effective
TPS cross sections as a function of the experimentally determined $\sigmaeff$ parameter. 
From~\cite{dEnterriaSnigirev}
\begin{eqnarray} 
\int d^2 b~ T^2_{\rm AA}({\bf b})= \frac{\rm A^4 (n+1)^2}{4\pi R^2_A (2n+1)}=\rm A^2\,T_{\rm AA}(0)\,\frac{n+1}{2n+1}\nonumber
\label{h-s,T2}
\end{eqnarray}
one obtains, via Eqs.~(\ref{eq:sigmaeffTPS})--(\ref{eq:sigmaeff1}), the 
relationship between $\sigmaefftps$ and $\sigmaeff$:
\begin{eqnarray} 
\sigmaefftps^2 = \frac{\rm A^2 (3n+1)(n+1)}{(2n+1)^2}\cdot\sigmaeff^2\,,
\label{h-s,T3-T2}
\end{eqnarray}
which for a proton-proton collision reads:
\begin{eqnarray} 
\sigmaefftps^2 =\frac{(3n+1)(n+1)}{(2n+1)^2}\cdot\sigmaeff^2\,.
\label{h-s,sigma3-2}
\end{eqnarray}
For the distributions given by (\ref{h-s,T}) with exponents $n= 0-\infty$, one obtains
$\sigmaefftps = [1-0.87]\times\sigmaeff$, \ie, the effective
TPS and DPS cross sections are numerically very similar. This result holds
for more realistic hadron profiles. 
Indeed, modern \pp\ Monte Carlo (MC) event generators, such as \pythia~\cite{Sjostrand:2007gs},
often parametrize the \pp\ overlap function in the form:
\begin{eqnarray} 
T({\bf b})= \frac{m}{2\pi r^2_p \Gamma (2/m)} \exp{[-(b/r_p)^m]} \,, 
\label{e,T}
\end{eqnarray}
normalized to one, 
where $\Gamma (2/m)$ is the gamma function.
The exponent $m$ depends on the MC ``tune'' obtained from fits to the measured \pp\ ``underlying event''
and various DPS cross sections~\cite{Khachatryan:2015pea}. It varies between a pure Gaussian ($m=2$) 
to more peaked exponential-like ($m=0.7, 1$) distributions. From the corresponding 
integrals of the square and cube of $T({\bf b})$:
\begin{eqnarray} 
& &\int d^2 b~ T^2({\bf b})= \frac{m}{2\pi r^2_p \Gamma (2/m) 2^{2/m}}\,,
\label{e,T2}
\end{eqnarray}
and
\begin{eqnarray} 
\int d^2 b~ T^3({\bf b})= \left[\frac{m}{2\pi r^2_p \Gamma (2/m)}\right]^2 \frac{1}{3^{2/m}},
\label{e,T3}
\end{eqnarray}
we obtain, via Eqs.~(\ref{eq:sigmaeffTPS})--(\ref{eq:sigmaeff1}),
\begin{eqnarray} 
  \sigmaefftps =(3/4)^{1/m}\cdot\sigmaeff\,,
\label{e,sigma3-2}
\end{eqnarray}
which is again independent of the exact numerical value of the proton ``radius'' $r_p$,
but depends on the overall shape of its transverse profile characterized by the exponent $m$.
For typical \pythia\ $m= 0.7,1,2$ exponents tuned from experimental data~\cite{Khachatryan:2015pea}, one obtains 
$\sigmaefftps~=~[0.66,0.75,0.87]\times\sigmaeff$
respectively. 
An alternative phenomenologically-motivated description of the proton profile is
given by the dipole fit of the two-gluon form factor in the momentum representation~\cite{Blok:2010ge}
\begin{eqnarray} 
F_{2g}({\bf q})=1/(q^2/m^2_g+1)^2,
\label{dip}
\end{eqnarray}
where the gluon mass $m_g$ parameter characterizes its transverse momentum $q$ distribution, 
and the transverse density is obtained from its Fourier-transform:
$f({\bf b})=\int e^{-i{\bf b}\cdot{\bf q}}F_{2g}({\bf q})\frac{d^2q}{(2\pi)^2}$.
Such a proton profile is used in other standard \pp\ MCs, such as \herwig~\cite{Seymour:2013qka}.
The corresponding DPS and TPS effective cross sections read~\cite{Snigirev:2016uaq}:
\begin{equation} 
\label{eq:sigmaeffdip}
\sigmaeff= \left[\int  F_{2g}^4(q)\frac{d^2q}{(2\pi)^2}\right]^{-1}=\frac{28 \pi}{m^2_g} , 
\end{equation}
and $\sigmaefftps^2 =$
\begin{eqnarray} 
& & \bigg[\int (2\pi)^2 \delta({\bf q_1}+{\bf q_2}+{\bf q_3}) F_{2g}({\bf q_1}) F_{2g}({\bf q_2}) F_{2g}({\bf q_3}) \nonumber \\
& & \times F_{2g}({\bf -q_1}) F_{2g}({\bf -q_2}) F_{2g}({\bf- q_3}) \frac{d^2q_1}{(2\pi)^2}\frac{d^2q_2}{(2\pi)^2}\frac{d^2q_3}{(2\pi)^2}\bigg]^{-1}.\nonumber
\end{eqnarray}
Numerically integrating the latter and combining it with (\ref{eq:sigmaeffdip}), we 
obtain $\sigmaefftps~=~0.83\times\sigmaeff$, quite close to the value obtained for the Gaussian \pp\ overlap function.

We note that in order to reproduce the experimentally measured $\sigmaeff \simeq 15$~mb, 
one should fix the characteristic proton ``size'' to $r_p \simeq 0.11,0.24,0.49$~fm for 
\pp\ overlaps of the form given by Eq.~(\ref{e,T}) with $m=0.7,1,2$; to $r_p \simeq 0.35, 0.40, 0.46$~fm 
for exponents $n= 0, 1, 2$ as defined in Eq.~(\ref{h-s,T}); and to $r_g=1/m_g \simeq 0.13$~fm in the case of
the two-gluon dipole fit (\ref{dip}). Despite the wide range of effective proton radius parameter derived, 
one of the main conclusion of this study is the robustness of the $\sigmaefftps\simeq\sigmaeff$ result. Indeed, 
the average and standard deviation of all typical parton transverse profiles considered here, yields
\begin{equation}
\label{eq:TPS_DPS_factor}
\sigmaefftps = k\times\sigmaeff, \; {\rm with}\;\; k = 0.82\pm 0.11\,,
\end{equation}
\ie, for the typical $\sigmaeff\simeq 15 \pm 5$ values extracted from a wide range of DPS measurements 
at Tevatron~\cite{Abe:1997xk} and LHC~\cite{Astalos:2015ivw,Abe:1997xk,Khachatryan:2015pea,Aaboud:2016dea,Aaij:2015wpa}, we finally obtain
\begin{equation}
\label{eq:TPS_factor}
\sigmaefftps = 12.5 \pm 4.5 \;{\rm mb}.
\end{equation}


The experimental observation of triple parton scatterings in \pp\ collisions requires perturbatively-calculable
processes with not too high energy (mass) scales so that their SPS cross section are not much smaller
than $\mathcal{O}(\rm 1\;\mu b)$ since, otherwise, their corresponding TPS cross section (which goes as the cube of 
the SPS values) are extremely reduced. 
Indeed, according to Eq.~(\ref{doubleAB}) with the data-driven estimate (\ref{eq:TPS_factor}), a triple hard process 
$\rm \pp\to a\,a\,a$, with SPS cross sections $\sigmaSPS_{pp\to a}\approx \rm 1\;\mu b$, 
has a very small TPS cross section $\rm \sigma^{\rm TPS}_{pp \to a\,a\,a}\approx 1$~fb, already without
accounting for extra reducing factors from decay branching ratios and experimental acceptances and reconstruction inefficiencies.
Promising processes to probe TPS, with not too small pQCD cross sections, are 
inclusive charm $\pp\to\ccbar+X$, and bottom $\pp\to\bbbar+X$ production.
These processes are dominated by gluon-gluon fusion, $gg\to\QQbar$, and at high energies receive 
contributions from scatterings at very small $x$, 
for which one can expect a non-negligible probability of DPS~\cite{Luszczak:2011zp,Berezhnoy:2012xq,Cazaroto:2013fua} 
and TPS in their total inclusive production. 

\begin{figure*}[htpb!]
\centering
\includegraphics[width=0.99\columnwidth]{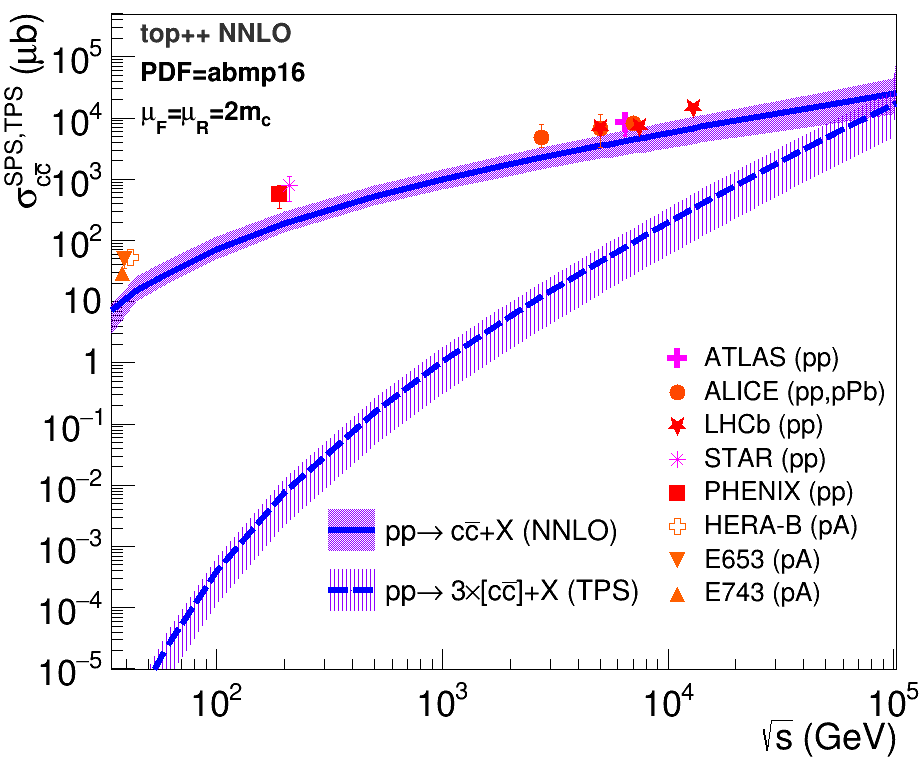}
\includegraphics[width=0.99\columnwidth]{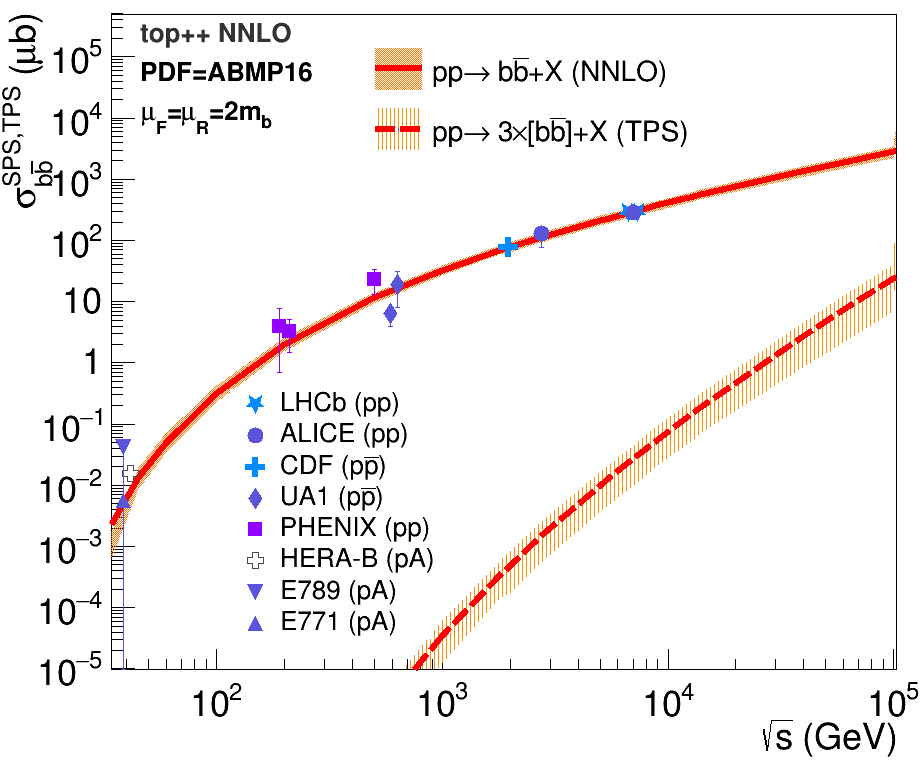}
\caption{Total charm (left) and bottom (right) cross sections in \pp\ collisions as a function of \cm\ energy, 
in single-parton (solid line) and triple-parton 
(dashed line) parton scatterings. Bands around curves indicate scale, PDF (and $\sigmaefftps$, in the case of $\sigmaTPS$) uncertainties added in quadrature.
The symbols are experimental data collected in~\cite{DdE}.
\label{fig:1}}
\end{figure*}

The TPS heavy-quark cross sections are computed via Eq.~(\ref{doubleAB}) for $\mathpzc{m}=1$, \ie\
$\sigma_{pp \to \QQbar}^{\rm TPS} = (\sigma_{pp \to \QQbar}^{\rm SPS})^3/(6\sigmaefftps^2)$
with $\sigmaefftps$ given by (\ref{eq:TPS_factor}), and $\sigma_{pp \to \QQbar}^{\rm SPS}$ 
calculated via Eq.~(\ref{eq:hardS}) at NNLO accuracy
using a modified version~\cite{DdE} of the $\toppp$ (v2.0) code~\cite{Czakon:2013goa}, with $\rm N_f=3,4$  
light flavors, heavy-quark pole masses at $\rm m_{c,b}=1.67, 4.66$~GeV, default renormalization
and factorization scales $\rm \mu_R=\mu_F=2\, m_{c,b}$, and using the ABMP16 PDFs~\cite{Alekhin:2016uxn}. 
The NNLO calculations increase the total SPS heavy-quark cross sections by up to 20\% at LHC energies
compared to NLO results~\cite{fonll,mnr}, reaching a better agreement with the experimental data, 
and featuring much reduced scale uncertainties ($\pm50\%,\pm15\%$ for $\ccbar$,$\bbbar$)~\cite{DdE}.
Figure~\ref{fig:1} shows the resulting total SPS and TPS cross sections for charm and bottom production
over $\sqrts =$~35~GeV--100~TeV, and Table~\ref{tab:1} collects a few values with associated uncertainties 
for nominal LHC and FCC energies.
The TPS cross sections are small but rise fast with $\sqrts$, as the cube of the SPS cross sections.
Although triple-$\bbbar$ cross sections remain quite small and reach only about 1\% of the inclusive bottom
cross section at the FCC ($\sqrts = 100$~TeV), triple-$\ccbar$ production from three independent parton 
scatterings amounts to 5\% of the inclusive charm yields at the LHC ($\sqrts = 14$~TeV) and to more than half
of the total charm cross section at the FCC. Since the total \pp\ inelastic cross section at $\sqrts = 100$~TeV
is $\sigma_{\rm pp}\simeq$~105~mb~\cite{dEnterria:2016oxo}, charm-anticharm triplets are expected to be produced
in about 15\% of the \pp\ collisions at these energies.
\renewcommand\arraystretch{1.2}%
\begin{table}[htpb]
\caption{\label{tab:1}
Total charm and bottom SPS (NNLO) and TPS cross sections (in mb) in \pp\ at LHC and FCC
with scales, PDF, and total (quadratic, including $\sigmaefftps$) uncertainties.}
\begin{ruledtabular}
\begin{tabular}{lccc}
 Final state  &  $\sqrt{s}=14$ TeV & $\sqrt{s}=100$ TeV \\ 
\colrule
$\sigma(\rm \ccbar+X)$ & $7.1\pm3.5_{\rm sc}\pm0.3_{\rm pdf}$ &  $25.0\pm16.0\pm1.3_{\rm pdf}$\\
$\sigma(\rm \ccbar\,\ccbar\,\ccbar+X)$ & $0.39\pm0.28_{\rm tot}$  &  $16.7\pm11.8_{\rm tot}$ \\
\colrule
$\sigma(\rm \bbbar+X)$ & $0.56\pm0.09_{\rm sc}\pm0.01_{\rm pdf}$ &  $2.8\pm0.6_{\rm sc}\pm0.1_{\rm pdf}$ \\
$\sigma(\rm \bbbar\,\bbbar\,\bbbar+X)$ & $(0.19\pm0.12_{\rm tot})\,10^{-3}$ & $(24\pm17_{\rm tot})\,10^{-3}$ \\
\end{tabular}
\end{ruledtabular}
\end{table}

The study presented here demonstrates that triple-parton scatterings are a non-negligible source of perturbative 
particle production in \pp\ collisions at increasingly higher energies. The formulas derived here
allow one to estimate the TPS yields for any final-state of interest. Using accurate NNLO predictions for single
heavy-quark production, we have shown that the production cross section of three $\ccbar$-pairs from three separate 
parton interactions is, in principle, observable at the LHC, and approaches the total charm cross section at 
$\sqrts \approx 100$~TeV.

{\it Acknowledgments--\;} Discussions with A.P.~Kryukov and M.A. Malyshev on TPS, and with 
M.~Cacciari, M.~Czakon, A.~Mitov and G.~Salam on NNLO heavy-quark calculations are 
gratefully acknowledged. 



\begin{references}
\bibitem{Diehl:2011yj}M.~Diehl, D.~Ostermeier, A.~Schafer, JHEP {\bf 03}(2012)089.
\bibitem{Bartalini:2011jp} P. Bartalini {\it et al.,} arXiv:1111.0469 [hep-ph].
\bibitem {Abramowicz:2013iva} H.~Abramowicz {\it et al.,}  arXiv:1306.5413 [hep-ph].
\bibitem{Bansal:2014paa} S.~Bansal {\it et al.,} arXiv:1410.6664 [hep-ph].
\bibitem{Astalos:2015ivw} R~Astalos {\it et al.,} arXiv:1506.05829 [hep-ph].
\bibitem{dEnterria:2010xip} D.~d'Enterria {\it et al.}, 
   Eur.\ Phys.\ J.\ C {\bf 66} (2010) 173.

\bibitem{Calucci:2010wg} G.~Calucci, D.~Treleani, Phys.\ Rev.\ D {\bf 83} (2011) 016012.
\bibitem{CalucciTreleani}G.~Calucci and D.~Treleani, Phys. Rev. D {\bf 79}, 074013 (2009);
Phys. Rev. D {\bf 79}, 074013 (2009);
Phys. Rev. D {\bf 80}, 054025 (2009);
Phys. Rev. D {\bf 86}, 036003 (2012).
\bibitem{Maina:2009sj} E.~Maina, J. High Energy Phys. 09  (2009) 081.
\bibitem{Snigirev:2016uaq} A.M.~Snigirev, Phys. Rev. D {\bf 94}, 034026 (2016).
\bibitem{Mangano16b}M. Mangano, G. Zanderighi, {\it et al.}, CERN-TH-2016-112; arXiv:1607.01831 [hep-ph].

\bibitem{Abe:1997xk}CDF Collaboration, Phys. Rev. D {\bf 56} (1997) 3811. 
\bibitem{Khachatryan:2015pea}
 CMS Collaboration,  Eur.\ Phys.\ J.\ C {\bf 76} (2016) 155.
\bibitem{Aaboud:2016dea} ATLAS Collaboration, JHEP {\bf 11} (2016) 110.
\bibitem{Aaij:2015wpa}LHCb Collaboration, JHEP {\bf 07} (2016) 052
\bibitem{Seymour:2013qka}M.~H.~Seymour and A.~Siodmok,   JHEP {\bf 10} (2013) 113.

\bibitem{dEnterriaSnigirev}
D.~d'Enterria and A.~M. Snigirev, Phys. Lett. B718 (2013) 1395; 
Phys. Lett. B727 (2013) 157. 

\bibitem{Sjostrand:2007gs} T.~Sj\"ostrand, S.~Mrenna and P.~Z.~Skands,  Comput.\ Phys.\ Commun.\  {\bf 178} (2008) 852.
\bibitem{Blok:2010ge}B.~Blok, Yu.~Dokshitzer, L.~Frankfurt, and M.~Strikman, Phys. Rev. D {\bf 83}, 071501 (2011).
\bibitem{Luszczak:2011zp}M.~Luszczak, R.~Maciula, and A.~Szczurek, Phys. Rev. D {\bf 85} (2012) 094034. 
\bibitem{Berezhnoy:2012xq}A.V.~Berezhnoy {\it et al.}, 
Phys. Rev. D {\bf 86}(2012) 034017.
\bibitem{Cazaroto:2013fua}E.R.~Cazaroto, V.P.~Goncalves, and F.S.~Navarra, Phys. Rev. D {\bf 88} (2013)  034005.

\bibitem{DdE}David d'Enterria, to be submitted.
\bibitem{Czakon:2013goa}M.~Czakon, P.~Fiedler and A.~Mitov, Phys.\ Rev.\ Lett.\  {\bf 110} (2013) 252004.
\bibitem{Alekhin:2016uxn}S.~Alekhin {\it et al.}, 
arXiv:1609.03327 [hep-ph]. 
\bibitem{fonll} M.~Cacciari {\it et al.}, 
JHEP {\bf 10} (2012) 137. 
\bibitem{mnr}M.~L.~Mangano, P.~Nason and G.~Ridolfi, Nucl.\ Phys.\ B {\bf 373} (1992) 295.
\bibitem{dEnterria:2016oxo} D.~d'Enterria and T.~Pierog, JHEP {\bf 08} (2016) 170.

\end{references}
\end{document}